# High Field Anomalies of Equilibrium and Ultrafast Magnetism in Rare-Earth-Transition Metal Ferrimagnets

A. Pogrebna, [1] K. Prabhakara,[1] M. Davydova,[2] J. Becker,[1,3] A. Tsukamoto,[4] Th. Rasing,[1] A. Kirilyuk[1]
A. K. Zvezdin,[2] P. C. M. Christianen,[3] and A.V. Kimel[1]

[1] Radboud University, Institute of Molecules and Molecules, Nijmegen, The Netherlands

[2] Prokhorov General Physics Institute of the Russian Academy of Sciences, Moscow, Russia

[3] High Field Magnet Laboratory (HFML –EMFL), Radboud University, Nijmegen, The Netherlands

[4] College of Science and Technology, Nihon University, Chiba, Japan

**ABSTRACT:** Magneto-optical spectroscopy in fields up to 30 Tesla reveals anomalies in the equilibrium and ultrafast magnetic properties of the ferrimagnetic rare-earth-transition metal alloy TbFeCo. In particular, in the vicinity of the magnetization compensation temperature, each of the magnetizations of the antiferromagnetically coupled Tb and FeCo sublattices show triple hysteresis loops. Contrary to state-of-the-art theory, which explains such loops by sample inhomogeneities, here we show that they are an intrinsic property of the rare-earth ferrimagnets. Assuming that the rare-earth ions are paramagnetic and have a non-zero orbital momentum in the ground state and, therefore, a large magnetic anisotropy, we are able to reproduce the experimentally observed behavior in equilibrium. The same theory is also able to describe the experimentally observed critical slowdown of the spin dynamics in the vicinity of the magnetization compensation temperature, emphasizing the role played by the orbital momentum in static and ultrafast magnetism of ferrimagnets.

## Introduction

The need for ever faster and energy-efficient data storage and information processing is a strong motivation to search for unconventional ways to control magnetism by means other than magnetic fields [1–4]. Several successful realizations of magnetization control with the help of an electric current [5,6], electric field [2] and light [3] have been demonstrated. This has become heavily debated topic in modern magnetism and revealed a lack of understanding of the mechanisms that are responsible for these phenomena [3,7]. It is clear, however, that in all these cases the spin-orbit and the exchange spin-spin interactions play a decisive role. For instance, spin-orbit-torques [8–12], multiferroicity [13,14], opto- and photomagnetic [3] phenomena cannot be understood without taking into account the spin-orbit interaction as well as orbital momenta of the ground and excited states. The exchange interaction, in turn, can be efficiently harnessed for spin manipulation in multi-sublattice spin systems or multilayers [15–20].

Ferrimagnetic rare-earth intermetallics, and rare-earth-transition metal (RE- TM) alloys in particular, are among the most studied systems in fundamental and applied magnetism. For example, unique functionalities of GdFeCo, GdFe, and GdCo alloys have been demonstrated in spintronics [21,22] and ultrafast magnetism [23–25]. The interplay between the exchange and the spin-orbit interaction in rare-earth ferrimagnets facilitate electric field, current and optical control of spins. In particular, due to the antiferromagnetic coupling between the non-equivalent magnetic sublattices in GdFeCo, it is possible to reverse its magnetization solely with a single femtosecond laser pulse [26]. Anomalous hysteresis loops and critical slowing down of laser-induced spin dynamics in high magnetic fields were reported for GdFeCo [27], but a theory of such behavior has not been developed until now. Interestingly, as the Gd ion in the alloy is in an *S*-ground state, its substitution with other rare-earth ions having non-zero orbital momentum in the ground state must enhance the spin-orbit interaction and thus open up fundamentally new opportunities in the field of spintronics and ultrafast magnetism.

TbFeCo is an example of such a material, that, due to large coercive fields above 10 T, is well suitable for high density magnetic recording [28,29]. Although several attempts of modeling the laser-induced spin dynamics in TbFeCo have been performed [30–32], not only spin dynamics, but even the static spin structure of unperturbed TbFeCo are poorly understood. Moreover, experimental studies of high-coercive-field materials are seriously hampered by the need for even higher magnetic fields and thus require unique experimental installations.

Here we report the experimental observation of anomalous hysteresis loops of the magnetizations of the antiferromagnetically coupled FeCo and Tb sublattices of ferrimagnetic TbFeCo in high magnetic fields. The loops appear to be very sensitive to temperature near the magnetization compensation point, where the magnetizations of the two sublattices cancel each other. We show that the observed behavior can be explained in the framework of an *f-d* ferrimagnet by taking into account the orbital momentum and, as a result, the large magnetic anisotropy of the rare-earth ions. In order to bring the developed theory into an ultimate test, we experimentally investigate the magnetization dynamics in TbFeCo triggered by a femtosecond laser pulse in high magnetic fields and compare the outcome of the experiment with the modeling. It is surprising that the theory developed to explain equilibrium magnetic properties is also able to predict the experimentally observed critical slowing down of the spin dynamics that was observed in the experiment.

The paper is organized as follows. Section I describes static magneto-optical measurements in which anomalous hysteresis loops were observed. Next, in section II we introduce the theoretical model aimed to describe the ground state of a ferrimagnet with large rare-earth anisotropy. We plot magnetic field-temperature (*H-T*) phase diagram and explain the equilibrium magnetism of TbFeCo using the developed theory. In section III we present the experimental results on laser-induced ultrafast dynamics of TbFeCo in high magnetic field, and theoretically explain the observed anomalies using the derived phase diagram. We conclude the manuscript with a summary, which emphasizes the simplicity, and at the same time predictive power of the proposed theory. Finally, we highlight experimentally observed features whose explanation, being beyond the capabilities of the presented model, is the next challenge in the physics of rare-earth alloys.

## I. Magneto-optical spectroscopy in high magnetic fields

The studied material was an amorphous rare-earth-transition metal alloy with stoichiometric composition $Tb_{22}Fe_{68.2}Co_{9.8}$. Without applied magnetic field, the antiferromagnetic exchange coupling between the rare-earth and the transition metal magnetic sublattices favor a collinear antiparallel alignment. The magnetizations of the transition metal and the rare-earth sublattices have different temperature dependences so that at the magnetization compensation temperature $T_M$ the net magnetization is zero, if no magnetic field is applied. In the studied sample $T_M$=322 K. Below this temperature, $T < T_M$, the net magnetization is dominated by the RE sublattice. Above the compensation point, the situation is the opposite, and the TM sublattice dominates the net magnetization. The strong inter-sublattice $3d$-$4f$ exchange interaction between the Tb and FeCo magnetic moments defines the Curie temperature, $T_C$, which is around 700 K [33]. The studied sample is a thin film structure with composition glass/SiN(5nm)/RE-TM(20nm)/SiN(60nm). The TbFeCo magnetic layer has an easy-axis type of magnetic anisotropy, oriented perpendicularly to the sample plane.

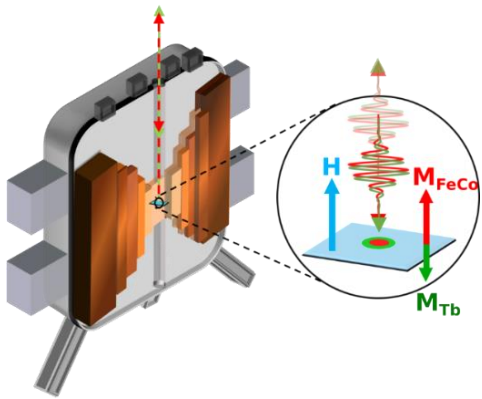

**Fig. 1 (color online)** *Schematic of the experiment. The sample was inserted into 37 T dissipative magnet. Dashed lines – pump and probe beams.*

The experiments were performed at the High Field Magnet Laboratory (HFML) in Nijmegen. Magnetic fields up to 30 T were applied along the normal to the sample (see Fig. 1). To benefit from elemental specificity of the magneto-optical phenomena in TbFeCo in the visible spectral range [34], we performed the measurements of the polar magneto-optical Kerr effect (MOKE) with light of two photon energies. In particular, for photon energy $\hbar\omega$=1.55 eV the effect is expected to probe the normal component of the magnetization of the FeCo sublattice, while light with $\hbar\omega$=2.41 eV is mainly sensitive to the normal magnetization component of Tb.

Figure 2 shows the MOKE as a function of the applied magnetic field at different temperatures near the magnetization compensation temperature. Above $T_M$ = 322 K the observed behavior is anomalous as the loops have a triple shape and are very sensitive to the sample temperature. At temperatures below $T_M$ = 322 K, the anomalies are far less pronounced, but still visible (see arrows in Fig. 2)

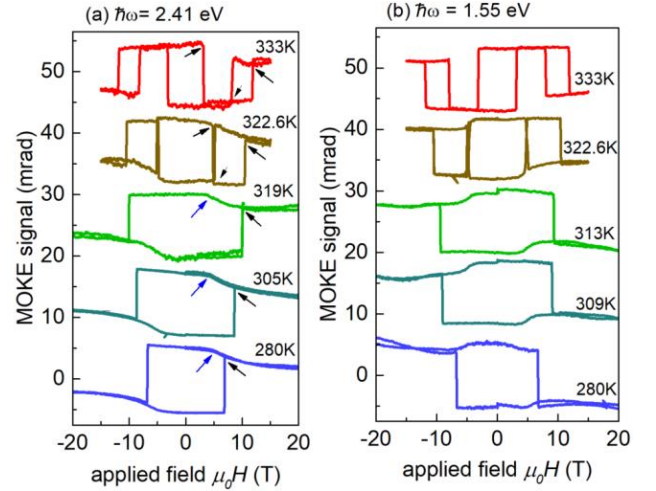

**Fig. 2 (color online)** *Static MOKE data of the TbFeCo sample measured at 2.41 eV (a) and 1.55 eV (b) photon energies at different temperatures from 267.0 K to 342.5 K. A paramagnetic background was subtracted from the measurements. The magnetization compensation temperature $T_M$ is around 322 K. Black arrows correspond to the hysteresis edges around the first-order field-induced phase transitions discussed in section II, while a second-order phase transition is shown with a blue arrow.*

Similar hysteresis loops were observed in rare-earth ferrimagnetic alloys and intermetallics earlier and explained by inhomogeneities with the strongest ever reported exchange bias fields [35,36]. In particular, an application of that theory to our case would imply that the exchange bias between the inhomogeneities reaches 10 T. However, it is also known that hysteresis loop dependencies of the magnetization on temperature or field is a signature of a first order phase transition. Gradual changes of the magnetization upon a change of temperature or field are generally explained as second-order phase transitions. Despite intense interest in rare-earth- transition metal alloys, their equilibrium properties in high magnetic fields and $H$-$T$ phase diagrams near the compensation temperature, in particular, are still unexplored. $H$-$T$ phase diagram of uniaxial ferrimagnets has been studied theoretically earlier [37], under the assumption that only the TM-sublattice is anisotropic. In systems with different symmetry [38,39] it was found that the behavior of the phase diagram is greatly affected by magnetic anisotropies of both sublattices.

In order to reveal the origin of the observed anomalous hysteresis loops, we developed a theory of magnetism of TbFeCo in thermodynamic equilibrium and calculated the *H-T* phase diagrams of this compound.

Magnetic fields at which magnetization changes were subtracted from the static MOKE data (see Fig. 2) in a wide range of temperatures are shown in phase diagram in Fig. 3(a). The blue line data points correspond to a second order phase transition, where the magnetization changes gradually. Black diamonds and associated black dashed lines correspond to the edges of the experimental hysteresis, where the magnetization changes abruptly.

## II. H-T phase diagram and anomalous hysteresis loops

Similarly to Refs [37,40], the calculations are based on analysis of the free energy for a two-sublattice *f-d* (RE-TM) ferrimagnet assuming that rare-earth ion is paramagnetic. We take advantage of the fact that the single-domain model works especially well in the vicinity of the compensation point. This is because the domain size diverges at the compensation point [41] as the magnetostatic energy drops to zero when the magnetization $M = M_f - M_d$ vanishes at $T_M$. The free energy has the form:

$$W = \mathbf{M}_d \cdot \mathbf{H} - \int_0^{H_{eff}} M_f(h,T)dh + W_a(\mathbf{M}_d, \mathbf{M}_f), \quad (1)$$

where $\mathbf{H}_{eff} = \mathbf{H} - \lambda \mathbf{M}_d$ is the effective magnetic field acting upon the rare-earth ion, $W_a = K_d \frac{(\mathbf{M}_d \cdot \mathbf{n})^2}{M_d^2} + K_f \frac{(\mathbf{M}_f \cdot \mathbf{n})^2}{M_f^2}$ is the uniaxial anisotropy energy, with $K_d, K_f$ being *d*- and *f*-sublattice anisotropy constants, respectively, and $\mathbf{n}$ is an easy axis unit vector. The magnetization function for rare-earth ions was taken in the following form:

$$M_f(H_{eff}, T) = \chi_1(T)H_{eff} + \chi_2(T)H_{eff}^3 + \chi_3(T)H_{eff}^5, \quad (2)$$

and it is directed along the effective field. Whereas for Gd the *f*-sublattice magnetization is accurately described by the Brillouin function, it doesn't hold well for other rare-earth ions with non-zero orbital momentum in the ground state [42], which is further complicated by the amorphous nature of the alloy. Therefore, we fit the susceptibilities numerically, starting from the functions obtained by an expansion of the Langevin function $L(\frac{\mu_{Tb}^{eff} H_{eff}}{k_B T})$ in series and numerically adapting them and the effective magnetic parameters to reproduce the experimentally observed features. We present the calculations performed for the following parameters: $H_{ex} = \lambda M_d \approx 80$ T, $T_M$ = 322 K, $T_C$ = 700 K [33]. Magnetic moments of the Tb and FeCo sublattices used in the calculations are $\mu_{Tb}^{eff} = 10\mu_B$ and $\mu_{FeCo}^{eff} = 1.8\mu_B$, respectively [31]. As it was shown earlier [38], the rare-earth anisotropy is one of the defining factors for the character of the phase diagram near the compensation point. In our model we assume that $K_f \approx 3 \times 10^6$ erg/cc and $K_d = 0$ (we assume the *d*-sublattice anisotropy to be much smaller than that of the rare-earth sublattice, as Fe ions are in the *S*-ground state).

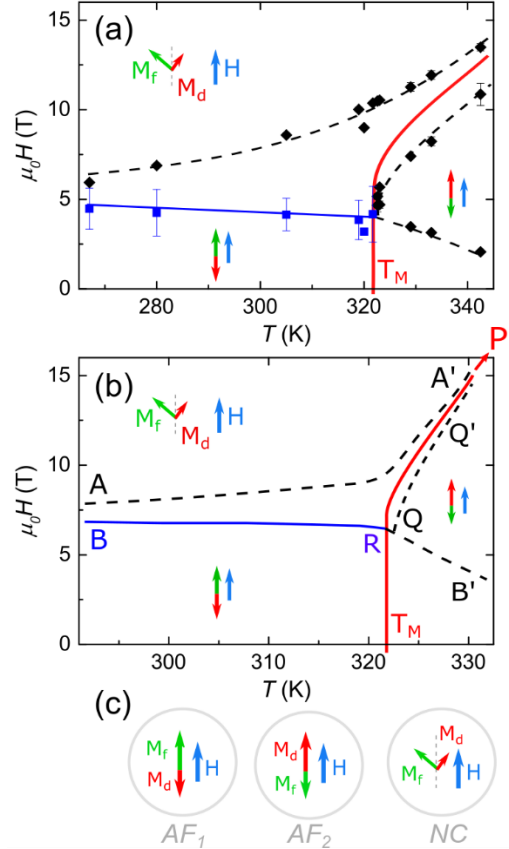

**Fig. 3 (color online)** (*a*) *Experimental and* (*b*) *theoretical magnetic field - temperature phase diagram for TbFeCo. The lines in the experimental phase diagram are guides for the eyes. The black points (dashed black curves) correspond to the points stability loss, the red curve ($T_M$P) is the first-order transition line, and the blue points (blue curve BR) correspond to the second-order phase transition.* (*c*) *Schematics of the three phases AF$_1$, AF$_2$ and NC that are present in the diagram. The states of the magnetization of both sublattices are shown with arrows, where the red arrow corresponds to the FeCo magnetization, and the green arrow corresponds to the Tb magnetization.*

Using expression (1) for the thermodynamic potential we numerically calculate the magnetic (*H-T*) phase diagram. The ground states of the system are found by minimization of the thermodynamic potential with regard to the order parameter $\theta_d$, which denotes the polar angle of the FeCo magnetization in the coordinate system with the *z*-axis aligned along the external magnetic field. At the local minima one finds $\frac{\partial \Phi}{\partial \theta_d} = 0$ and $\frac{\partial^2 \Phi}{\partial \theta_d^2} > 0$. The lines of stability loss, where $\frac{\partial^2 \Phi}{\partial \theta_d^2} = 0$, are found for each phase. At the lines of the first-order phase transition two phases, corresponding to solutions $\theta_d^{(1)}$ and $\theta_d^{(2)}$, correspond to the local minima of the thermodynamic potential and the condition $\Phi\left(\theta_d^{(1)}\right) = \Phi\left(\theta_d^{(2)}\right)$ is fulfilled.

The phase diagram in Fig. 3(a,b) shows three phases: two antiferromagnetic collinear phases $AF_1$ and $AF_2$ (rare-earth magnetization is along the magnetic field below the compensation temperature $T_M$ in phase $AF_1$, and opposite to it above $T_M$ in phase $AF_2$) and one canted phase $NC$. The blue line $BR$ represents the second-order phase transition between phases $AF_1$ and $NC$ and defines the spin-flop field denoted as $H_{BR}$ below the compensation point. Above the compensation temperature the spin-flop occurs discontinuously, via a first-order phase transition (see line $RP$ in Fig. 3 (a)). The red line ($T_MP$) corresponds to a first-order phase transition between the collinear phases $AF_1$ and $AF_2$ along segment $T_MR$ as well as between the phases $AF_2$ and $NC$ along the rest of the line, i.e. along segment $RP$. Each of the dashed curves corresponds to the stability loss of one of the phases. Following the markup in Fig. 3 (a), lines $AA'$, $QQ'$ and $RB'$ (we denote the corresponding fields $H_{AA'}$, $H_{QQ'}$ and $H_{RB'}$, respectively) are the lines of stability loss of phases $AF_2$, $NC$ and $AF_1$, respectively. One might notice that the first-order phase transition $T_MRP$ goes to the outside of the area shown in the phase diagram. The point $P$ is a tricritical point at which the order of the phase transition between collinear and noncollinear phases changes from first to second. A number of unusual phenomena are expected to occur in ferrimagnets near this point [43,44], being an interesting subject for future studies.

The features of the magnetic phase diagram can be observed experimentally by measuring the dependence of the magnetization on external magnetic field. In particular, deducing magnetic fields corresponding to jumps in the experimentally measured hysteresis loops, we define the fields and the temperatures at which collinear and non-collinear phases loose stability. In this way we plotted the experimental phase diagram shown in Fig. 3(a). The structure of the theoretical phase diagram (see Fig. 3(b)) explains well the behavior of the experimentally observed phase transitions. To demonstrate the agreement of the theoretical predictions with the experimental results, we also calculate the dependence of all possible equilibrium values of the order parameter as a function of external magnetic field $\theta_d^{(i)}(H)$ at fixed temperature. The obtained magnetization plots explain the experimental results shown in Fig. 2. Above the compensation temperature, as illustrated in Fig. 4 (a), the theory reproduces the triple hysteresis loops observed in the experiment. The large central loop at lower fields encompasses the first-order phase transition between the two collinear phases. Two loops that appear at higher positive and negative fields are due to the two first-order phase transitions between collinear and non-collinear phases (from $AF_2$ to $NC$ and vice versa, respectively). From Fig. 3 (a) one can see that the first-order transition, from which the additional loop at positive field originates, occurs at the line $RP$. The hysteresis around these first-order phase transitions is defined by the position of the stability loss lines in the ($H$-$T$) phase diagram and corresponds to the dashed lines in Fig. 3 (a). The loops disappear if the phase transition to the non-collinear phase $NC$ becomes of second-order. At 290 K, i.e. below the compensation point, we see that the second-order phase transition occurs below the coercive field (see Fig. 4(b)).

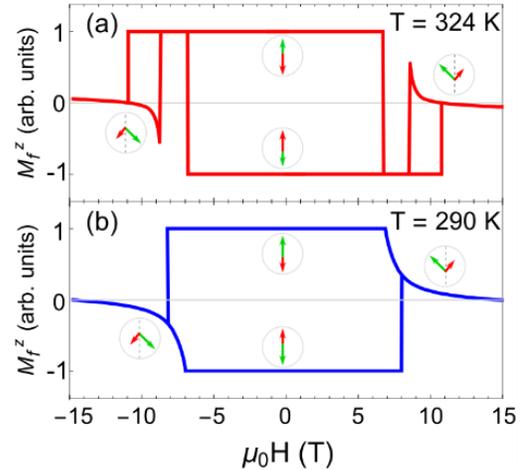

**Fig. 4 (color online)** *Calculated Tb magnetization curves (a) below and (b) above compensation temperature, at T = 290 K and T = 324 K, respectively.*

Therefore, our relatively simple theory is able to qualitatively explain the observed anomalous hysteresis loops without involving inhomogeneities and huge exchange bias fields. The single-domain picture holds well near the compensation point, where the domain size in the magnet diverges. The triple hysteresis loops can be explained as a series of first-order phase transitions. The quantitative difference between the theory and the experiment can be improved by taking into account such features of realistic amorphous alloys as random single ion anisotropy, resulting in sperimagnetism [45].

To test our theory further, it is interesting to check if the theory can also explain anomalies in ultrafast magnetism of rare-earth-transition metal ferrimagnets. Some of these anomalies were seen in earlier experimental studies of ultrafast laser-induced spins dynamics in GdFeCo in the vicinity of the spin-flop transition [29], but neither theory nor simulations of the dicovered high field dynamics have been reported up to now.

### III. Ultrafast magnetism and critical slow-down in high magnetic fields

To investigate the dependence of ultrafast spin dynamics on bias temperature and high magnetic field, we performed time-resolved measurements of the polar magneto-optical Kerr effect (tr-MOKE) using a pump-probe technique, with 50 fs optical pulses generated by a 1-kHz Ti:Al2O3 regenerative amplifier seeded with a Ti:Al2O3 oscillator. The central photon energy of the pulses could be tuned with the help of an Optical Parametric Amplifier. Relying on the conventionally accepted approximation that the main effect of optical pump pulse on a metallic magnet is ultrafast heating and relying on conclusions of earlier studies [33], we assumed that the laser-induced spin dynamics is independent of the photon energy of the pump. Tuning the photon energy of probe one can be sensitive to FeCo and Tb sublattices [34], respectively. Therefore, we performed two types of pump-probe experiments. In order to monitor the laser-induced spin dynamics of the Tb-sublattice, the pump and probe photon energies were chosen at 1.55 eV and 2.48 eV, respectively. In order to monitor the dynamics of the FeCo-sublattice, the pump and probe photon energies were altered. The pump beam was focused on the sample into a spot around 90 µm in diameter and the diameter of the probe beam was smaller - around 30 µm. The fluence of the pump pulses was 0.15 mJ/cm², while the probe fluence was kept around 1.5 µJ/cm². All time-resolved measurements were performed at magnetic fields outside the hysteresis loops.

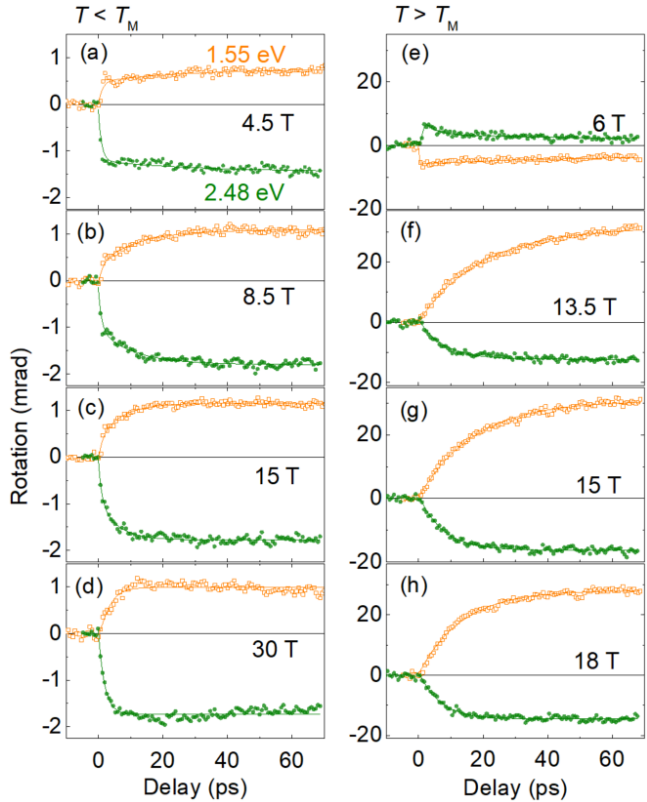

**Fig. 5 (color online)** *Transient magneto-optical Kerr effect measured from TbFeCo at different magnetic fields. Traces measured with probe photon energy at 1.55 eV (FeCo sublattice) are shown by open orange squares. Experiments performed at photon energy 2.48 eV (Tb sublattice) are shown by filled green circles. In the left panel shown traces measured below the compensation point, T = 160 K (1.55 eV) and T = 220 K (2.48 eV). In the right panel shown traces measured above the compensation point, T = 350 K. The lines are corresponding fits with functions, discussed in the text.*

The results of tr-MOKE measurements on TbFeCo at various magnetic fields and at different probe photon energies below and above the compensation point are shown in left and right panels of Fig. 5, respectively.

To analyze the magnetization dynamics, we will distinguish two time-domains: (i) sub-10 ps longitudinal dynamics i.e. demagnetization of the magnetic sublattices and (ii) sub-100 ps transversal dynamics of the magnetic sublattices, i.e. tilt of the magnetization. The data shown in Fig. 5 were fitted with a double-exponential function $\sim A_0 \exp(-t/\tau_0) + A_1 \exp(-t/\tau_{rise})$, where $\tau_0$ and $\tau_{rise}$ are the characteristic times of the ultrafast longitudinal and transversal dynamics, respectively. $A_0$ and $A_1$ are the amplitudes. Assuming that the ultrafast demagnetization of both Tb and FeCo sublattices is completed within a few ps [33], in the fit we set $\tau_0$ = 1.5 ps, while $A_0$, $A_1$, $\tau_{rise}$ were taken as fitting parameters (see Supplementary II for details).

In the collinear phase (Fig. 5 (e)) the dynamics is very fast and associated with the longitudinal demagnetization (i), which is in good agreement with previous reports [27,33,46,47].

Figure 6 shows the field dependencies of the rise time $\tau_{rise}$, as deduced from the fit, below and above the compensation temperature in the noncollinear phase. It is clearly seen that the dynamics slows down close to the spin-flop transition. It is remarkable that below the compensation temperature, decreasing the external magnetic field from $H_{AA'}$+5 T to $H_{AA'}$, which is close to the spin-flop field $H_{BR}$ at that temperature, leads to a 400% increase of the rise time. A similar decrease of the field from $H_{AA'}$+5 T to $H_{AA'}$ above the compensation temperature results in a rather moderate increase of $\tau_{rise}$ by 25%. Finally, we find that above the compensation temperature the magnetization of Tb and Fe have clearly different dynamics with a slower response of the Tb spins.

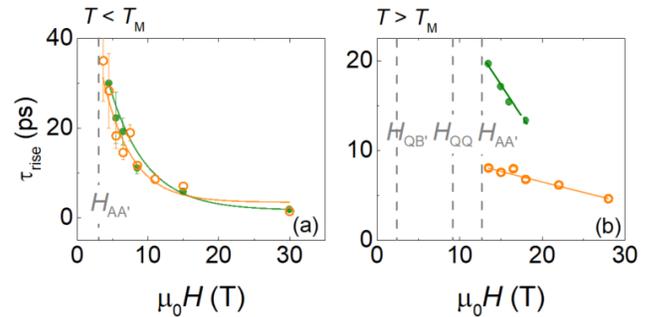

**Fig. 6 (color online)** *(a, b) The rise time of the tr-MOKE signal (see Fig. 5) below and above the compensation point, respectively. Orange circles (open) and curves show characteristic times constants which correspond to the FeCo magnetic sublattice, while green circles (filled) and curves correspond to the Tb magnetic sublattice. Grey dashed lines correspond to the hysteresis edge around the magnetic field-induced first order phase transitions.*

To assign the observed features of the presence and absence of the critical slowing down to the peculiarities of the phase diagram, we simulate the ultrafast laser-induced magnetization dynamics. We start with the corresponding magnetic structure in thermodynamic equilibrium and assume that a femtosecond laser pulse demagnetizes both sublattices by 10% (see Supplementary II). The subsequent transversal magnetization dynamics was modeled with the help of the Landau-Lifshitz-Gilbert (LLG) equation. We show that the observed dynamics in the canted phase can be explained in the framework of coherent magnetization precession of thermalized sublattices brought out of equilibrium by ultrafast demagnetization. After the demagnetization, the spins of the sublattices will relax towards the equilibrium orientation via a heavily damped precession.

In the framework of the LLG equation, one can derive the out-of-equilibrium position of the magnetization right after the demagnetization [48]. We divide the magnetization dynamics into three time-domains, as earlier: after the initial longitudinal demagnetization (i), the coherent rotation of spins further away from the initial equilibrium orientation occurs (ii). We find that a change in magnetization of any of the sublattices of the order of one percent is already enough to trigger the magnetization dynamics similar to that observed in the experiment. Using the framework described above we derived analytical expressions for the rise time, corresponding to the strongly damped dynamical regime as observed in the experiment:

$$\tau_{rise}^{pr} = \frac{\text{arctanh}\left[\left(\sqrt{(\alpha\omega_{ex}/2)^2 - \omega_r^2}\right)/(\alpha\omega_{ex}/2)\right]}{\left(\sqrt{(\alpha\omega_{ex}/2)^2 - \omega_r^2}\right)}, \quad (3)$$

where $\alpha$ is the effective Gilbert damping constant for the ferrimagnet, the exchange frequency $\omega_{ex} = \gamma H_{ex}$, and the resonance frequency is defined as $\omega_r^2 = \gamma^2\left[\frac{m}{\chi_\perp}H\cos\theta_0 + \left(\frac{2(K_f - K_d)}{\chi_\perp} - H^2\right)\cos 2\theta_0\right]$, where $m = M_d - M_f$ and $\gamma$ is the gyromagnetic ration of electron. The angle $\theta_0$ is the angle between the external magnetic field and the antiferromagnetic vector $L = M_d - M_f$ and $\chi_\perp$ is the component of the magnetic susceptibility perpendicular to the external magnetic field. We assume the effective Gilbert damping constant, which is a function of the composition and temperature [49], to be equal to 0.2. An example of the calculated dynamics can be found in Fig. S4 (see Supplementary).

The phase diagram predicts that below the compensation point the transition to the angular phase upon an increase of the external magnetic field occurs via a second-order phase transition (Fig. 3). At the phase transition the frequency of the ferromagnetic resonance softens, $\omega_r \to 0$, and the dynamics of the order parameter slows critically (diverges to infinity) down as predicted by equation (3) and calculations shown in Fig. 7 (a). Similar slowing down is seen experimentally (Fig. 6 (a)).

Figure 7 (b) summarizes the calculated field dependency of the response time above the compensation temperature, where the phase transition between collinear and angular phases is of first order. The slowing down at the phase transitions is not critical, as it would be expected at a first-order transition from the general theory of phase transitions [50]. Therefore, the experimental results reported in Fig. 7 (b) agree well with the theoretically predicted behavior based on the magnetic phase diagram. Note that experimental verification of the theoretically predicted features of first-order phase transitions is often obscured by such factors as sample inhomogeneities.

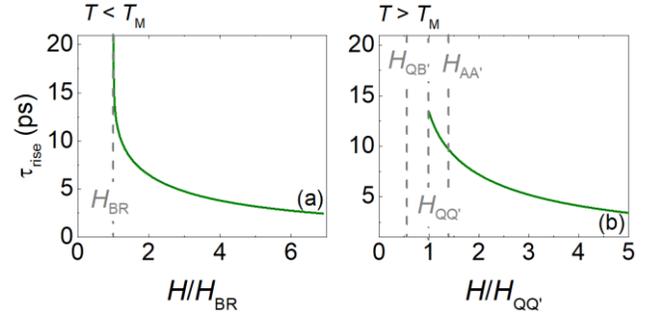

**Fig. 7 (color online)** *Dependence of the calculated rise time on the external magnetic field. The calculations were performed below (a) and above (b) the compensation temperature at $T_1 = 290$ K and $T_2 = 324$ K, respectively. The characteristic fields at the given temperatures are $H_{BR}(T_1) = 7$ T and $H_{QQ'}(T_2) = 8.7$ T.*

Finally, our model could not reproduce the dramatic difference between the timescales for the sublattices of Tb and FeCo observed experimentally (Fig. 7 (b)). We must note that a difference in dynamics of the two sublattices at the timescales of the order of 60 ps has been reported earlier [34], but despite several efforts of computational studies [19,51], the origin of such a behavior is still not understood. Different excitation times for TM and RE magnetic sublattices were previously observed in time-resolved X-ray studies [15]. The mechanism was explained by a larger magnetic moment per rare-earth ion in comparison to the magnetic moment of transition metal ions. However, reported experiments were done in the collinear phase where the rise time is on a 1 ps timescale, and the difference was observed only on timescales where the electron-phonon system is still not thermalized. Distinct spin dynamics on a time-scale 10-100 ps must have a different origin. We suggest that a possible explanation of such a behavior can be sperimagnetism reported for realistic TbFeCo alloys [45]. As a matter of fact, modeling spin dynamics of sperimagnets is one of the challenges in modern computational magnetism.

## Conclusions

We performed experimental and theoretical studies of anomalous hysteresis loops of the magnetizations of the antiferromagnetically coupled FeCo and Tb sublattices of ferrimagnetic TbFeCo in high magnetic fields. Unlike previous theories which explained such loops by exchange bias between the surface and bulk layers within one film, here we showed that such a loop can be an intrinsic feature a *f-d* ferrimagnet. By taking into account the orbital momentum that results in a large magnetic anisotropy of the rare-earth ions, we computationally explored and defined the phase diagram of TbFeCo in *H-T* coordinates. In order to bring the developed theory into an ultimate test, we experimentally investigated the magnetization dynamics in TbFeCo triggered by a femtosecond laser pulse and compared the outcome of the experiments with the modeling. It is surprising that the theory developed to explain equilibrium magnetic properties is also able to predict the experimentally observed dynamics, including critical slowing down of the order parameter in the vicinity of the magnetic compensation temperature. Finally, we note that above the compensation temperature, we experimentally observed clearly different dynamics of the magnetization of Tb and Fe sublattices. These features call for further theoretical studies that would take into account such features of realistic amorphous alloys as random single ion anisotropy and sperimagnetism.

## Acknowledgments

The work was funded by the Nederlandse Organisatie voor Wetenschappelijk Onderzoek (NWO), the European Research Council ERC Grant agreement No. 339813 (Exchange), HFML-RU/FOM member of the European Magnetic Field Laboratory (EMFL), the European Union Horizon 2020 and innovation program under the FET-Open grant agreement no.713481 (SPICE).